\documentclass{aastex63}
\usepackage{upgreek}

\newcommand{\degree}{$^{\circ}$}

\received{}
\revised{}
\accepted{}
\submitjournal{ApJ}

\shorttitle{Limiting Spectral Resolution -- EBL Reflection Grating}
\shortauthors{DeRoo et al. 2020}

\begin{document}

\title{Limiting Spectral Resolution of a Reflection Grating Made via Electron-Beam Lithography}

\correspondingauthor{Casey T. DeRoo}
\email{casey-deroo@uiowa.edu}

\author[0000-0002-9184-4561]{Casey T. DeRoo}
\author{Jared Termini}
\affiliation{The University of Iowa, \\
Dept. of Physics \& Astronomy, \\
Van Allen Hall, Iowa City, IA 52242, USA}

\author{Fabien Gris{\'e}}
\author{Randall L. McEntaffer}
\author{Benjamin D. Donovan}
\affiliation{The Pennsylvania State University, \\
Dept. of Astronomy \& Astrophysics, \
505 Davey Lab, \\
University Park, PA 16802, USA}

\author{Chad Eichfeld}
\affiliation{The Pennsylvania State University, \\
	Materials Research Institute, \
	N-153 Millennium Science Complex, \\
	University Park, PA 16802, USA}

\begin{abstract}
Gratings enable dispersive spectroscopy from the X-ray to the optical, and feature prominently in proposed flagships and SmallSats alike. The exacting performance requirements of these future missions necessitate assessing whether the present state-of-the-art in grating manufacture will limit spectrometer performance. In this work, we manufacture a 1.5 mm thick, 1000 nm period flat grating using electron-beam lithography (EBL), a promising lithographic technique for patterning gratings for future astronomical observatories. We assess the limiting spectral resolution of this grating by interferometrically measuring the diffracted wavefronts produced in $\pm$1st order. Our measurements show this grating has a performance of at least $R \sim $ 14,600, and that our assessment is bounded by the error of our interferometric measurement. The impact of EBL stitching error on grating performance is quantified, and a path to measuring the period error of customized, curved gratings is presented. 
\end{abstract}


\section{Introduction}
\label{sec:intro}
Encoded in spectra are the physics of astronomical sources. Temperature, density, relative motion, velocity distributions, and ionization states can all be deduced with sufficiently detailed spectra. Gratings are a critical component of dispersive spectrometers, which are employed across three orders of magnitude in wavelength space (1 -- 1000 nm). 

The dispersive spectrometers of future observatories are tasked with addressing critical science questions requiring high spectral resolution $R$ in order to untangle nearby line blends or detect faint features on top of a dominant background. For example, \emph{Lynx}, an X-ray Strategic Mission Concept under study for the 2020 Decadal Survey on Astronomy and Astrophysics (\cite{Gaskin_2019}), requires a dispersive grating spectrometer with $R > $ 5,000 in order to achieve its science objective of detecting the hot ($10^6$ -- $10^7$ K) filamentary structures thought to host much of the Universe's baryonic material at the current epoch (\cite{Bregman_2015} and references therein). Similarly, the Large Ultraviolet Optical Infrared Surveyor (LUVOIR), another Strategic Mission Concept, baselines two spectrometers operating in the ultraviolet with resolutions ranging from $R = $ 8,000 -- 65,000 (LUMOS, \cite{France_LUMOS_SPIE_2017}) to $R > $ 120,000 (POLLUX, \cite{Bouret_POLLUX_SPIE_2018}). These spectrometers address core science goals of LUVOIR, such as understanding the evolution of protoplanetary disks, probing the warm phases of the intergalactic medium through absorption spectroscopy, and characterizing the winds of metal-poor massive stars to understand their impact on feedback processes at early cosmic epochs. For these missions, large-format ($\gtrsim$ 10 cm$^2$) gratings are essential in order to capture a large portion of the light from the large-aperture telescopes.

The instrument optical designs for both \emph{Lynx} and \emph{LUVOIR} benefit from grating customization. For example, introducing a grating `chirp,' where the period is intentionally varied across the grating, theoretically increases the spectral resolution of the transmission grating spectrometer concept for \emph{Lynx} (\cite{Moritz_SPIE_2019}). Furthermore, the POLLUX instrument, a UV-spectropolarimeter baselined for \emph{LUVOIR}, employs gratings patterned on freeform optics i.e., optics with arbitrary deviations from a plane or curved surface (\cite{Muslimov_SPIE_2018}). These custom patterns must be realized over areas $\gtrsim$10 cm$^2$, and the resulting gratings blazed to offer high diffraction efficiency at high dispersion.  

In addition to enabling instruments for these flagship concepts, customized gratings also permit novel small missions with targeted science goals. Blazed gratings on curved surfaces would allow for further development of the two-element spectrometer concept for X-ray spectroscopy, a compact instrument that nonetheless would improve on the line detection sensitivity of the \emph{Chandra} and \emph{XMM-Newton} spectrometers by an order of magnitude (\cite{DeRoo_SPIE_CDXO_2019}). By manipulating both the local groove density and groove direction of the grating pattern, aberration-correcting gratings can be realized (\cite{Beasley_SPIE_2019}). Aberration-correcting gratings enable unique diffraction geometries, such as point-to-point imaging with only one optical element for a fixed wavelength or a hyperspectral imager operating from 400 -- 1000 nm compact enough for a CubeSat format (\cite{Beasley_SPIE_2016}). Finally, customized gratings can improve the performance of suborbital missions relying on efficiently blazed gratings (e.g., CHESS, \cite{Hoadley_SPIE_2016}; OGRE, \cite{Donovan_SPIE_2019}).  

The gratings for all of these missions must be inherently capable of greater spectral resolution than needed for the science case, as a realistic error budget for the spectrograph will degrade the resolution performance. Formally, the spectral resolution of a grating is limited by the total number of grooves $G$ and working order $n$ (\cite{Hutley_1982}), making the limiting spectral resolution of large-area gratings $R = nG \gtrsim 10^6$ in principle. In practice, however, fabrication errors dominate the achievable resolution. These fabrication errors are either related to the optical quality of the grating substrate, such as slope errors or microroughness, or errors in the grating pattern itself, such as deviations from the designed local grating period. 

\begin{figure}[h]
	\centering\includegraphics[width=3in,trim={0.0in 0.0in 0.0in 0.0in},clip]{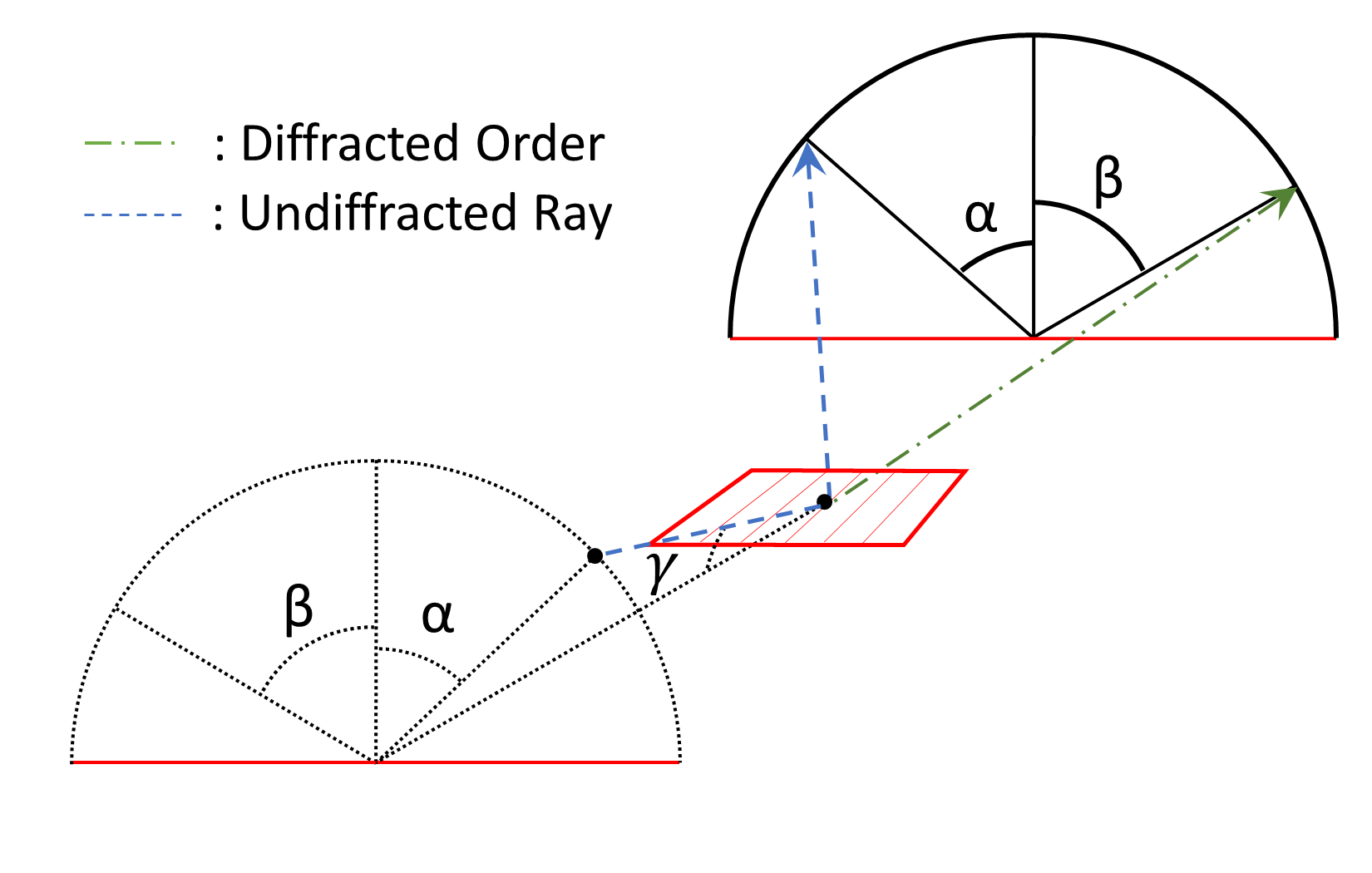}
	\caption{\label{fig:conical_diffraction} Cartoon of the generalized grating geometry. Light is incident on the grating at a cone angle $\gamma$, and reflects or diffracts to a point on a semi-circle with the same cone angle. The undiffracted light forms an angle $\alpha$ with the grating normal, whereas diffracted light travels to an angle $\beta$.}
\end{figure}

The deviations from the designed local grating period are related to the groove placement accuracy of the technique used to pattern the grating -- high groove placement accuracy yields a distribution of groove periods that closely match the grating's design, while poor groove placement accuracy results in unwanted variation in the space between adjacent grooves. These errors in groove placement enter as errors in the grating's period in the generalized grating equation (Fig. \ref{fig:conical_diffraction}):

\begin{equation}
\sin{\alpha} + \sin{\beta} = \frac{n\lambda}{d \sin{\gamma}},
\label{eq:grating_equation}
\end{equation}

\noindent as $\sigma_d$, the error in the grating groove period $d$. For a given incidence geometry (set by $\alpha$, the angle between the reflected beam and grating normal as projected into the focal plane, and $\gamma$, the cone angle between the center groove and the incidence light), an error in $d$ degrades the ability to measure the diffraction angle $\beta$ and hence the ability to determine the wavelength $\lambda$ at a given order $n$. Thus, an error in $d$ maps to an error in $\lambda$, limiting the resolution of a perfect spectrometer to $R = \lambda/\Delta\lambda \sim d/\sigma_d$. 

In sum, a host of astrophysics missions benefit from gratings fabricated with high groove placement accuracy and customized for the instrument application. Hence, a high-accuracy patterning technique flexible enough to produce (1) blazed (2) grating patterns with variable line spacing (VLS) over (3) large areas on (4) freeform optical surfaces is desirable. Several techniques promise a viable pathway towards meeting these requirements, including photolithography on freeform substrates (\cite{Muslimov_2018}) and mechanical ruling using atomic force microscopy (\cite{Gleason_SPIE_2017}). In this paper, we focus on grating manufacture via electron-beam lithography (EBL). EBL is a precise, flexible lithography method which rasters a beam of energetic, collimated electrons across resist. EBL can create minute feature sizes ($\sim$ 10 nm, \cite{Manfrinato_2013}), but has traditionally been limited to small areas ($\lesssim$ 1 cm$^2$). However, recent advances in EBL patterning have permitted the production of a VLS grating with a nominal $\sim$160 nm period over a 75 cm$^2$ area (\cite{Miles_2018}). EBL has also been used to write gratings with sculpted, triangular groove profiles directly using greyscale lithography and thermal reflow (\cite{McCoy_2018}). These direct-write blazed gratings have achieved high-diffraction efficiency in the X-ray when used in an echelle mounting (\cite{McCoy_2020}; \cite{McCurdy_SPIE_2019}). Finally, EBL has been previously employed to a write grating on a curved optical surface for the \emph{CRISM} instrument onboard the \emph{Mars Reconnaissance Orbiter}, albeit with a large period (15.552 $\upmu$m) and over a $\sim$ 4 cm$^2$ area (\cite{Wilson_SPIE_2003}). Thus, EBL has demonstrated large-formats, blazing, curved substrate patterning, and VLS pattering, making the method a viable technical path towards realizing gratings suitable for future astronomical instruments.

However, an assessment of the groove placement accuracy afforded by EBL for astronomical gratings is lacking. Measurements of the spectral resolving power of X-ray spectrometer systems under test offer lower limits on the period error (e.g., \cite{DeRoo_2020, Donovan_2020, Heilmann_2019}. However, these studies offer no spatial information about the achieved grating periodicity, and are routinely limited by the inherent width of the fluoresced X-ray lines or the focus quality of the employed telescope. Measuring the period error of EUV/X-ray gratings for use in synchrotrons has been previously done (e.g., \cite{Voronov_2017}; \cite{Gleason_SPIE_2017}); however, these gratings are on thick substrates not suitable for astronomical use given the constraints on packing geometry and instrument mass.

In this paper, we characterize the groove placement accuracy achieved on an EBL-patterned astronomical grating using an optical interferometer. Interferometric measurements offer information about the frequency content and spatial distribution of period errors and are assessed independently of a spectrometer system. Moreover, the grating in the present study is patterned using the same tooling used to produce the large-format VLS grating of \cite{Miles_2018} and the direct-write blazed gratings of \cite{McCurdy_SPIE_2019}. Thus, these measurements characterize the present state-of-the-art for making high-resolution, customized gratings with EBL. A description of the grating under test and the interferometric measurements conducted are described in Section \ref{sec:methods}. In Section \ref{sec:results}, the groove placement accuracy and the derived groove period error are presented, along with a method of verifying the internal consistency of our measurements and an assessment of the noise inherent in our interferometric measurement. Finally, a discussion of the implications of these results for astrophysical instruments and a description of an interferometric technique for assessing EBL patterns on curved substrates is given in Section \ref{sec:discussion}.  

\section{Experimental Method}
\label{sec:methods}

\subsection{Measuring Gratings Interferometrically}
\label{subsec:interferometer_method}

\begin{figure}[h]
	\centering\includegraphics[width=4in,trim={0.0in 0.0in 0.0in 0.0in},clip]{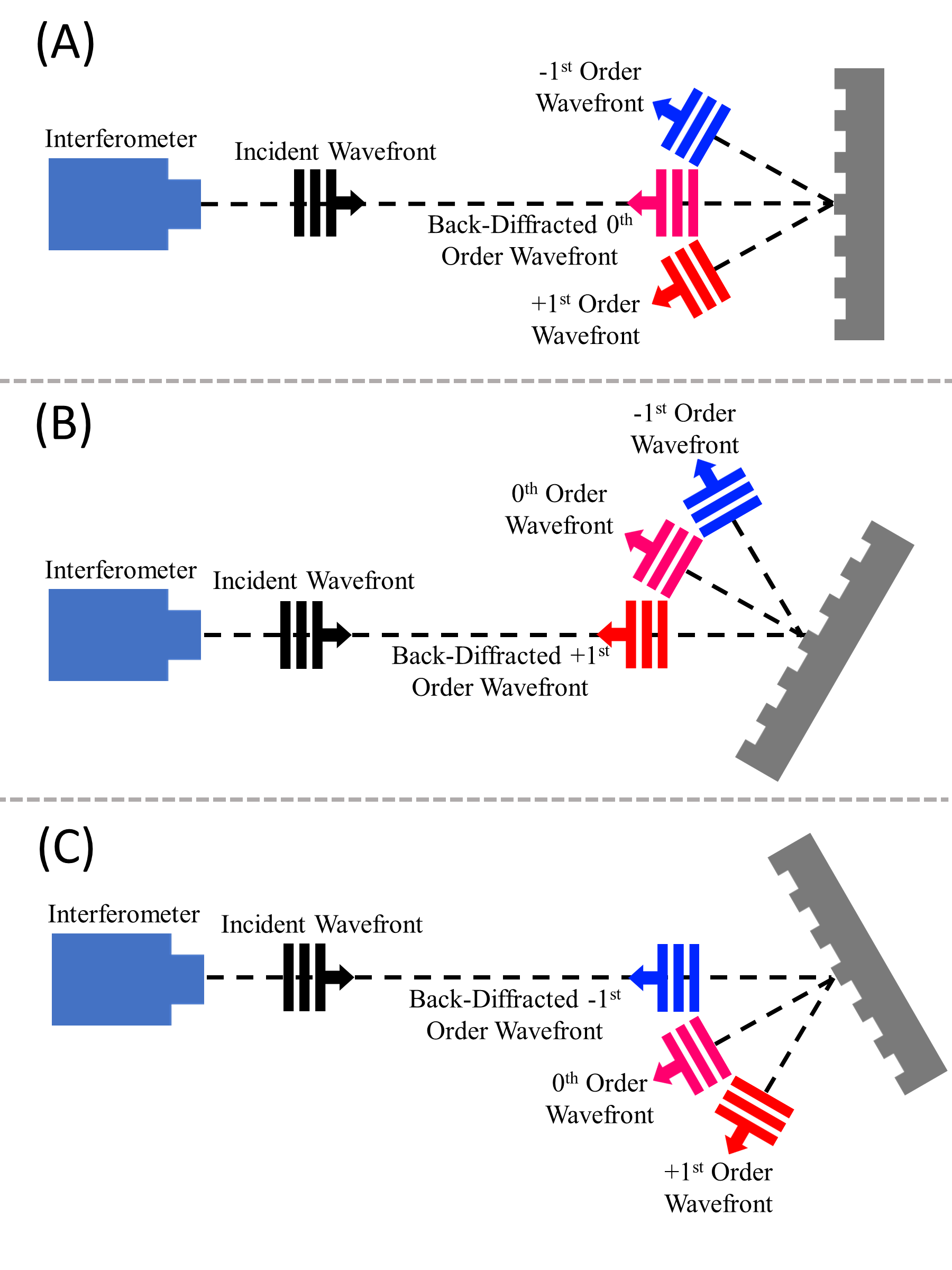}
	\caption{\label{fig:interferometric_backdiffraction} Geometry for the interferometric grating measurement. (A) The incidence geometry for measuring the 0th order of the grating. In this geometry, the grating behaves like a mirror and the figure of the optic is measured directly. (B) Rotation about the groove direction (out of the page) such that Eq. \ref{eq:backdiffraction} is satisfied aligns +1st order with the incidence wavefront, permitting the interferometer measurement of the back-diffracted wavefront. (C) Similar to (B), the opposite rotation permits the measurement of -1st order interferometrically. }
\end{figure}

Both the figure of a planar, constant-period grating and the optical quality of the diffracted orders can be assessed interferometrically, provided the constraints can be satisfied (\cite{Hutley_1982}). Measuring the grating's figure is straightforward -- in reflection, the grating behaves like a mirror. Thus, the optical figure of the grating can be measured with an interferometer equipped with a transmission flat by placing the grating at normal incidence to the beam (Fig. \ref{fig:interferometric_backdiffraction}A). A commercial interferometer integrates over the produced fringe patterns to yield a height map over the surface of the grating. In this paper, this measurement of the grating figure is referred to as the 0th order height map $\mathcal{H}_{0}(x,y)$, where $x$ is the coordinate in the dispersion direction and $y$ is along the groove direction. 

To measure the wavefront produced by the diffracted orders, the grating is placed into a geometry in which a diffracted order propagates back along the incident beam. This is achieved by setting $\gamma = $ 90\degree\,such that diffraction happens entirely in the plane of incidence, and rotating the grating about the groove direction such that the condition

\begin{equation}
\sin{\alpha} = \frac{n\lambda}{2d},
\label{eq:backdiffraction}
\end{equation}

\noindent is realized. This condition stems from evaluating Eq. \ref{eq:grating_equation} when $\beta = \alpha$. In this geometry, a coherent, back-diffracted wavefront produced by the grating can be measured by the interferometer. As a result of projection effects, this wavefront is diminished in lateral extent by a factor of $\cos{\alpha}$ relative to the 0th order height map. The height map for diffracted order $n$ is referred to as $\mathcal{H}_{n}(x \cos{\alpha_n},y)$, where $\alpha_n$ is the angle $\alpha$ for which Eq. \ref{eq:backdiffraction} is satisfied for a given order $n$. 

An ideal grating will produce a back-diffracted wavefront that is perfectly in-phase -- the phase difference across the grating from the non-zero incidence angle are exactly offset by the phase difference induced by the grating pattern. Phase errors in the diffracted wavefront hence have one of two sources: (1) the figure error of the grating viewed in projection results in path length errors and thus phase errors between different physical areas of the grating or (2) local deviations from the average period change the local angle of diffraction, resulting in a difference in path length (and hence phase) when traversing the interferometric cavity. For a real (i.e., non-idealized) grating, both of these sources of error are present simultaneously. These errors are discussed in greater detail in \cite{Gleason_SPIE_2017}; we summarize key aspects of their work as pertains to interferometric grating measurement in the following paragraphs.

For grating figure, phase differences are proportional to $\mathcal{H}_{0}(x,y)$. At small incidence angles, these errors are symmetric about order i.e., identical for back-diffracted wavefronts with the same value of $|n|$. In the latter case of period error, grooves offset from their position in an idealized grating fail to cancel the phase difference introduced by placing the grating at a non-zero incidence angle. We define this offset in idealized groove position as $\Delta(x,y)$ for a given trace in the dispersion direction. $\Delta(x,y)$ quantifies the groove placement accuracy of a grating patterning technique (see Sec. \ref{sec:intro}), as it measures the offset between the realized and ideal groove positions. The phase errors introduced due to $\Delta(x,y)$ are antisymmetric about order since grooves are offset closer to or farther away from the source wavefront at opposite incidence angles, producing the opposite phase shift for $\pm$$n$. 

This difference in symmetry can be exploited to isolate these sources of error, yielding either the groove offset $\Delta(x,y)$ or the figure error $\mathcal{H}_{0}(x,y)$ from diffracted measurements of opposite orders $\mathcal{H}_{\pm\,n}(x,y)$. The height map measured by a commercial interferometer is related to the spatially-varying phase map $\phi(x,y)$ produced by the grating: 

\begin{equation}
\phi(x,y) = \frac{4\pi}{\lambda}\mathcal{H}_{n}(x,y),  
\label{eq:phase_height_relationship}
\end{equation}

\noindent \emph{modulo} a constant overall phase offset. Following the definitions of \cite{Gleason_SPIE_2017} and using Eq. \ref{eq:phase_height_relationship}, it can be shown that:

\begin{equation}
\mathcal{H}_{0}(x,y) = \frac{ \mathcal{H}_{+n}(x \cos{\alpha_n},y) + \mathcal{H}_{-n}(x \cos{\alpha_n},y)}{2 \cos{\alpha_n}},
\label{eq:figure_from_orders}
\end{equation}

\begin{equation}
\Delta(x,y) = \frac{d}{\lambda} \big( \mathcal{H}_{+n}(x \cos{\alpha_n},y) - \mathcal{H}_{-n}(x \cos{\alpha_n},y) \big),
\label{eq:displacement_from_orders}
\end{equation}

\noindent where $d$ is the average period of the grating under measurement and $\lambda$ the operating wavelength of the interferometer.

\begin{figure}[h]
	\centering\includegraphics[width=4in,trim={0.0in 0.0in 0.0in 0.0in},clip]{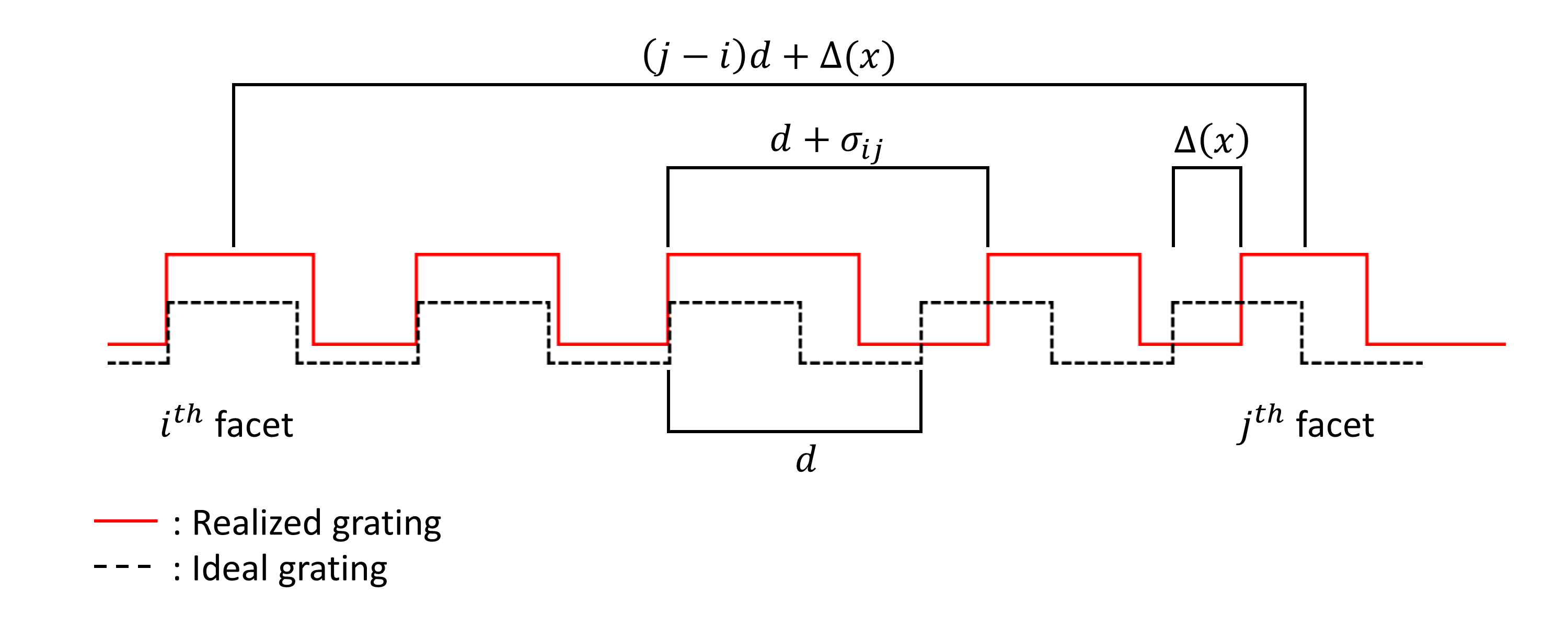}
	\caption{\label{fig:groove_offset_figure} A diagram depicting the differences between an ideal grating (black, dashed line) and the as-manufactured grating (red, solid line). The distance between the \emph{i}th and \emph{j}th facet is equal to the number of periods between them $(j - i)d$ plus a groove offset $\Delta(x)$. This groove offset is in turn related to the average period error by Eqs. \ref{eq:period_error_sum} and \ref{eq:average_period_error_1d}.}  
\end{figure}

Next, we relate the measured groove offset $\Delta(x,y)$ to period error as a function of position on the grating $\sigma_d(x,y)$ by considering the origin of a groove offset in one dimension. We define the distance between two grating grooves (\emph{i, j}) in the dispersion direction as $x_{i,j}$, which can be written as:

\begin{equation}
x_{i,j} = (j - i)d + \Delta(x_{i,j})  
\end{equation}
\noindent where $d$ is the average period of the grating, $(j - i)$ is the total number of grooves separating the two under examination and $\Delta({x_{i,j}})$ encompasses any remaining offset i.e., any deviation from an ideal grating.  $\Delta(x_{i,j})$, in turn, is the sum of the period errors for all the grooves between the \emph{i}th and \emph{j}th (see Fig. \ref{fig:groove_offset_figure}): 

\begin{equation}
\Delta(x_{i,j}) = \sum_{k = i}^{j} \sigma_k.
\label{eq:period_error_sum}
\end{equation}

\noindent where $\sigma_k$ is the error in the period of the \emph{k}th groove. 

To calculate a localized average period error $\sigma_d(x)$ as a function of position on the grating, we next make two assumptions. First, we assume that the spatial sampling in the dispersion direction $dx$ constitutes many periods i.e. $(j - i) >> 1$ such that $dx/d >> 1$. Next, we assume that the period error for each groove is small relative to the overall period of the grating, $\sigma_k << d$, such that $dx$ can be faithfully approximated as $(j -i)d$. Under these assumptions, Eq. \ref{eq:period_error_sum} can be converted to an integral and differentiated to yield:

\begin{equation}
\sigma_d(x) = d \frac{d\Delta(x)}{dx}
\label{eq:average_period_error_1d}
\end{equation}

\noindent  where $\sigma_d(x)$ represents the localized average period error. Eq. \ref{eq:average_period_error_1d} has a straightforward interpretation -- since the error in groove position accumulates at a rate of $\sigma_d(x)$ per groove, the error $d\Delta(x)$ over a spatial interval $dx$ is simply the number of grooves contained in the interval $(dx/d)$ multiplied by $\sigma_d(x)$. 

This expression is easily generalized to two dimensions by considering that the groove offset $\Delta(x,y)$ relative to an arbitrary reference point is determined wholly by the average period error accumulating in the dispersion direction $x$. Eq. \ref{eq:displacement_from_orders} and Eq. \ref{eq:average_period_error_1d} thus provide a means to move from a set of interferometric measuremements of diffracted orders $\mathcal{H}_{\pm\,n}(x \cos{\alpha_n},y)$ to a map of the average period error over the surface of the grating $\sigma_d(x,y)$.   

\subsection{EBL-Written Grating Measurements}
\label{subsec:intro_measurements}

\begin{figure}[h]
	\centering\includegraphics[width=3in,trim={0.0in 0.0in 0.0in 0.0in},clip,angle = 90]{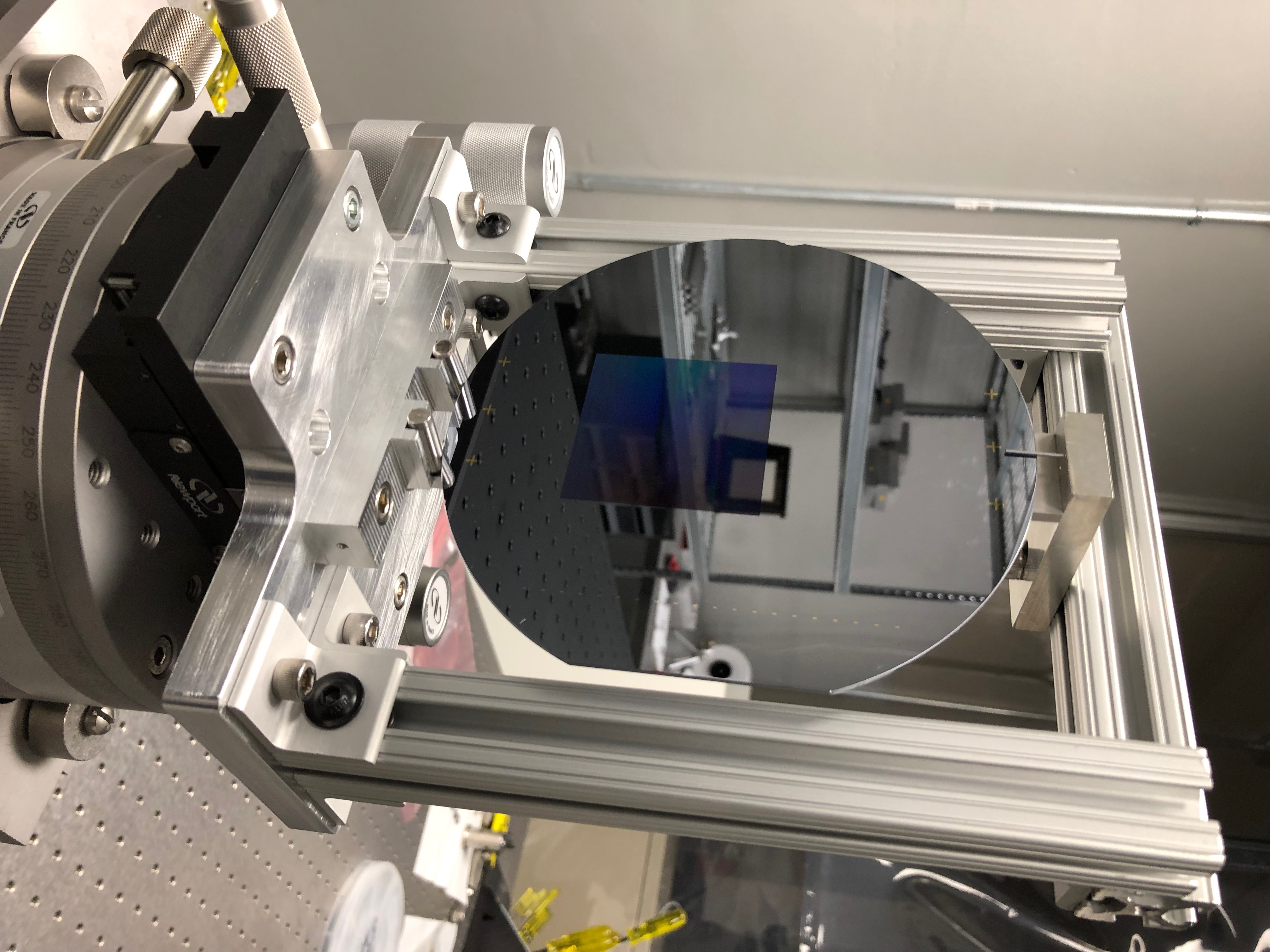}
	\caption{\label{fig:grating_picture} The 50 mm $\times$ 50 mm laminar grating written using EBL measured in this study. The grating area is visible in the lower center of the 6 in. diameter Si wafer as a darker, iridescent square. The grating substrate is held in a three-point mount, which is in turn affixed to rotation stages permitting alignment to the interferometer.}
\end{figure}

To assess the groove placement accuracy of astronomical gratings made using EBL, we manufactured a 50 mm $\times$ 50 mm, 1000 nm period grating with laminar grooves (Fig. \ref{fig:grating_picture}). The grating substrate is a 6 inch diameter, 1.5 mm thick silicon wafer. A 30 nm silicon nitride layer was deposited onto the silicon wafer with low-pressure chemical vapor deposition and served as a hard mask for the transfer of the pattern. This wafer was next coated with ZEP520A (1:1 dilution in anisole) and patterned at The Pennsylvania State University's Material Research Institute using a Raith EBPG5200 EBL tool. Following development, the pattern was transferred into the silicon substrate using a plasma etch performed by a Plasma-therm Versalock tool. The process for manufacturing astronomical gratings via EBL is described in greater detail in \cite{Miles_2018}.

For the present work, we measure both the grating's figure (0th order) and wavefront of the +1st and -1st diffracted orders in back-diffraction. With respect to Eq. \ref{eq:backdiffraction}, the back-diffaction angle $\alpha_{\pm1}$ for these measurements is 18.44\degree. Measurements were taken with a 4D Technologies AccuFizH100S Fizeau interferometer operated in dynamic mode in order to minimize the impact of environmental vibration during integrations (\cite{Brock_SPIE_DynamicInteferometry_2005}). This tool is equipped with a 6MP camera for ultra-fine sampling; the effective pixel scale of measurements operated in this mode is 0.043 mm/pix in 0th order.  

\begin{figure}[h]
	\centering\includegraphics[width=3in,trim={0.0in 0.0in 0.0in 0.0in},clip]{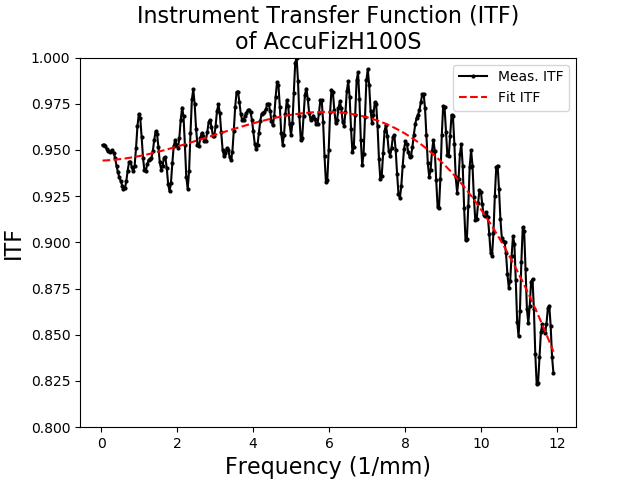}
	\caption{\label{fig:itf} The ITF of the AccuFizH100S Fizeau interferometer employed to measure the EBL grating produced for this study. The measured ITF is shown as a solid line, while the best-fit to the ITF is shown as dashed line.}
\end{figure}

From a Fourier optics perspective, the measured power in spatial errors may be reduced by the frequency response of the interferometer optical system (\cite{Goodman_IntroductionToFourierOptics_1996}; \cite{deGroot_ITF_2005}). To quantify this loss and support an analysis of the frequency content of the measured period error, the instrument transfer function (ITF) on the interferometer was independently measured. The ITF is the ratio of the power of a test optic measured by an optical system to the known power present on that test optic as a function of spatial frequency. Hence, the ITF is a measure of how accurately an optical system captures errors at a specific spatial frequency. The ITF can be applied as a corrective factor to provide a more accurate estimate of the spatial frequency content of an optic with unknown spatial frequency content. The ITF of the interferometer system employed was characterized prior to the grating measurements reported here, and is shown in Fig. \ref{fig:itf}. 

The measurement error of the wavefronts is estimated using the repeatability of the interferometer. The grating figure (0th order) was measured in separate instances, between which the grating was realigned to the interferometer. We assume that this repeatability error is comparable to the error present in our measurements at separate orders, as each order also requires a distinct alignment to the interferometer beam.  This error is used to assess the error present in our calculation of 0th order from $\pm$1st orders (see Sec. \ref{subsec:zero_from_first}) and in the average period error of the grating (see Sec. \ref{subsec:displacement_period_error}).

\section{Results}
\label{sec:results}

\begin{figure}[h]
	\centering\includegraphics[width=\textwidth,trim={0.0in 0.0in 0.0in 0.0in},clip]{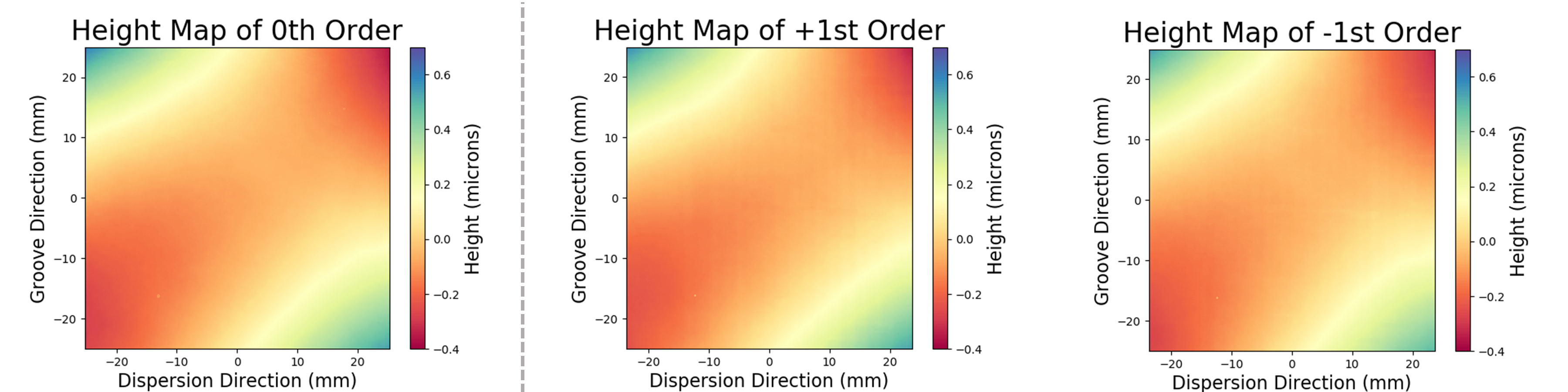}
	\caption{\label{fig:measured_grating_orders} The measured height maps of 0th order (left), +1st order (middle), and -1st order (right). All height maps are in microns, have the same color scale, and are plotted over the 50 mm $\times$ 50 mm extent of the grating. }
\end{figure}

The measured height maps of 0th, +1st, and -1st orders ($\mathcal{H}_{0}(x,y), \mathcal{H}_{+1}(x,y), \mathcal{H}_{-1}(x,y)$ respectively) are shown in Fig. \ref{fig:measured_grating_orders}. These height maps are normalized such that the average value across the height map is zero (i.e., that piston is zero) so they can be compared directly. These three measurements, along with the interferometer repeatability used as a proxy for the error implicit in these height maps, form the basis for the analysis presented in Sections \ref{sec:results} and \ref{sec:discussion}. 

\subsection{Methodology Verification -- Predicting Figure from the Diffracted Wavefronts}
\label{subsec:zero_from_first}
\begin{figure}[h]
	\centering\includegraphics[width=\textwidth,trim={0.0in 0.0in 0.0in 0.0in},clip]{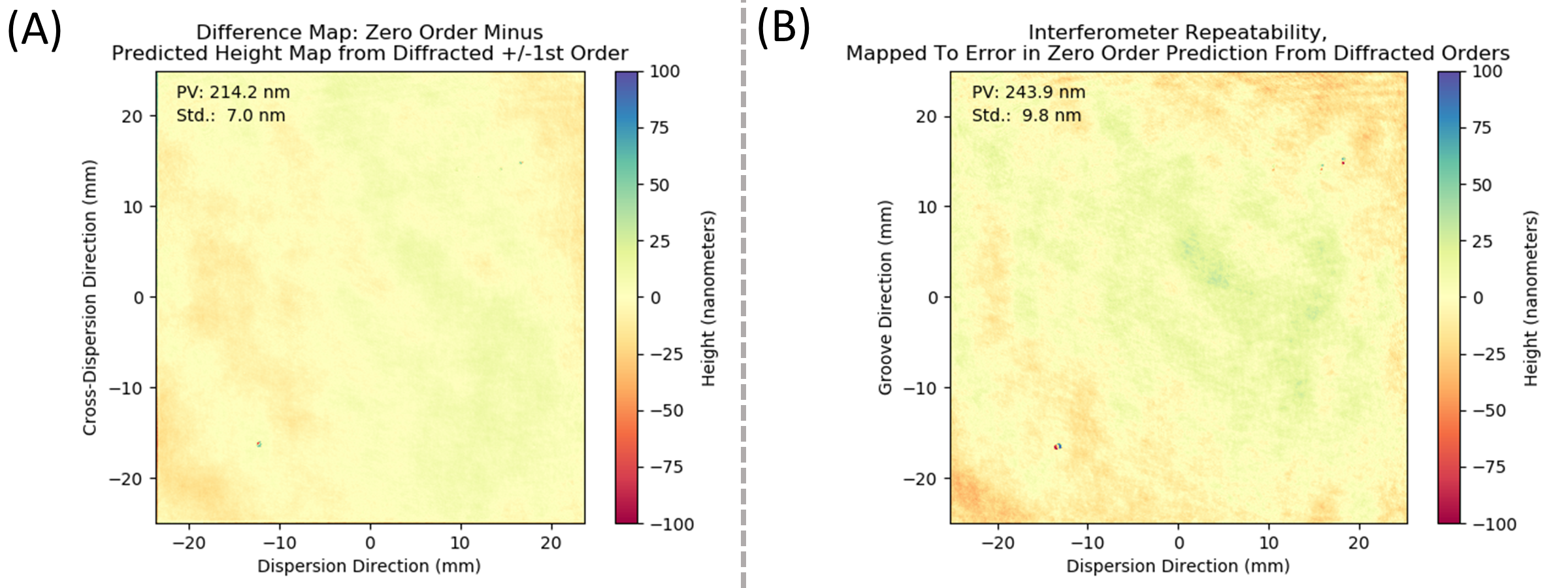}
	\caption{\label{fig:height_map_comparison} (A) The difference between the predicted grating figure (0th order) as calculated from $\mathcal{H}_{+1}(x,y)$ and $\mathcal{H}_{-1}(x,y)$ and the measured 0th order. This height map is equivalent to evaluating the lefthand side of Eq. \ref{eq:figure_from_orders_specific}. (B) The error in the difference map, as calculated by propagating the interferometer repeatability via Eq. \ref{eq:figure_from_orders_specific}. By either a peak-to-valley (PV) or RMS metric, the difference map agrees with zero to within measurement error. }
\end{figure}

As a cross-check on the self-consistency of the height map measurements, we use the height maps of the +1st and -1st orders to predict the figure of the grating as measured by the 0th order height map. From Eq. \ref{eq:figure_from_orders}, we see that:

\begin{equation}
\frac{ \mathcal{H}_{+1}(x \alpha_{\pm1},y) + \mathcal{H}_{-1}(x \cos{\alpha_{\pm1}},y)}{2 \cos{\alpha_{\pm1}}} - \mathcal{H}_{0}(x,y) = 0,
\label{eq:figure_from_orders_specific}
\end{equation}

\noindent where zero on the right hand side of the expression is interpreted in a statistical sense i.e., is consistent within the error of the interferometer. The left side of Fig. \ref{fig:height_map_comparison} displays the result of evaluating the left hand side of Eq. \ref{eq:figure_from_orders_specific} with the measured data. We find that this difference map is centered about zero as expected. Moreover, we find a characteristic variation of 7.0 nm RMS across the grating surface, consistent with the variation expected due to interferometer measurement noise (9.8 nm RMS). Thus, we conclude that our measurements are consistent with Eq. \ref{eq:figure_from_orders_specific}, bolstering confidence in the application of the methodology outlined in Section \ref{subsec:interferometer_method}.
 
\subsection{Groove Displacement and Groove Period Error}
\label{subsec:displacement_period_error}
\begin{figure}[h]
	\centering\includegraphics[width=\textwidth,trim={0.0in 0.0in 0.0in 0.0in},clip]{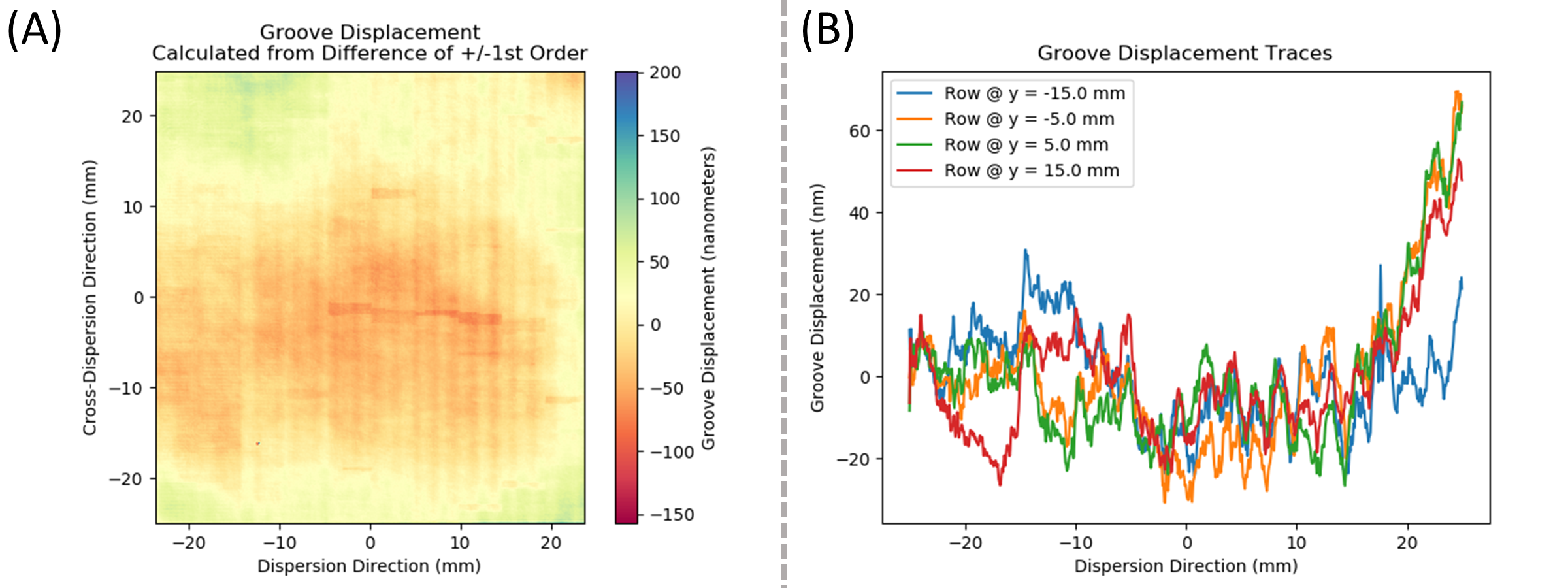}
	\caption{\label{fig:groove_displacement_map} (A) A map of the groove displacement $\Delta(x,y)$ over the grating surface as calculated from  $\mathcal{H}_{+1}(x,y)$ and $\mathcal{H}_{-1}(x,y)$ via Eq. \ref{eq:displacement_from_orders}. (B) Traces across the map of $\Delta(x,y)$ taken at fixed $y$ positions.}
\end{figure}

We next compute the position-dependent groove displacement $\Delta(x,y)$ from the height maps of the diffracted orders  using Eq. \ref{eq:displacement_from_orders}. The result is shown in Fig. \ref{fig:groove_displacement_map}. The left side of Fig. \ref{fig:groove_displacement_map} is the displacement map over the whole grating surface, $\Delta(x,y)$, while the right side shows several representative traces of groove displacement across the dispersion direction (i.e., $\Delta(x,y = c)$, where $c$ is a chosen constant). Note that we have remapped the pixel scale of these $\pm$1st order measurements to account for the $\cos{\alpha_{\pm1}}$ reduction in spatial sampling. In doing so, they can be compared directly to the 0th order height map (Fig. \ref{fig:measured_grating_orders}), albeit with coarser spatial sampling (0.045 mm/pixel vs. 0.043 mm/pixel). Periodic structure in this displacement map on the scale of a couple of millimeters is evident; an analysis of the power spectral density (PSD) of this measurement is performed in Sec. \ref{subsec:period_frequency}. 

\begin{figure}[h]
	\centering\includegraphics[width=\textwidth,trim={0.0in 0.0in 0.0in 0.0in},clip]{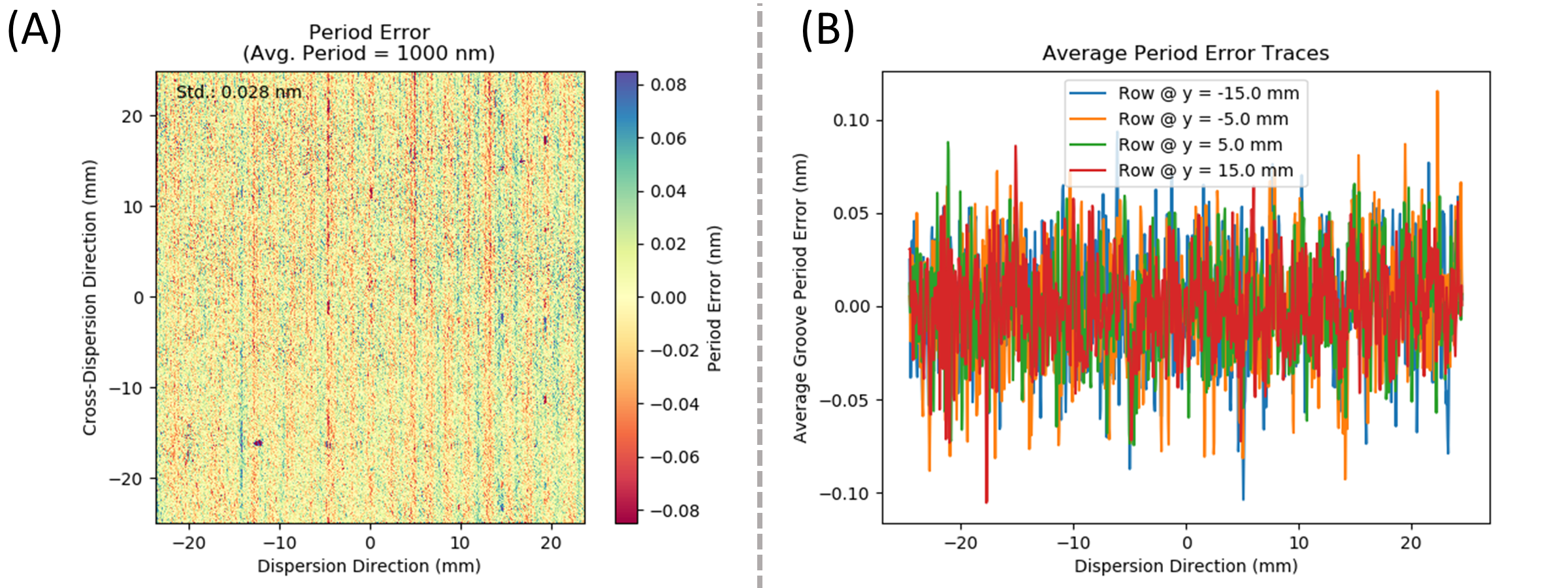}
	\caption{\label{fig:groove_period_map} (A) The average period error across the grating surface as calculated from Fig. \ref{fig:groove_displacement_map} and Eq. \ref{eq:average_period_error_1d}. (B) Traces across the average period error map at fixed $y$ positions.}
\end{figure}

\begin{figure}[h]
	\centering\includegraphics[width=3in,trim={0.0in 0.0in 0.0in 0.0in},clip]{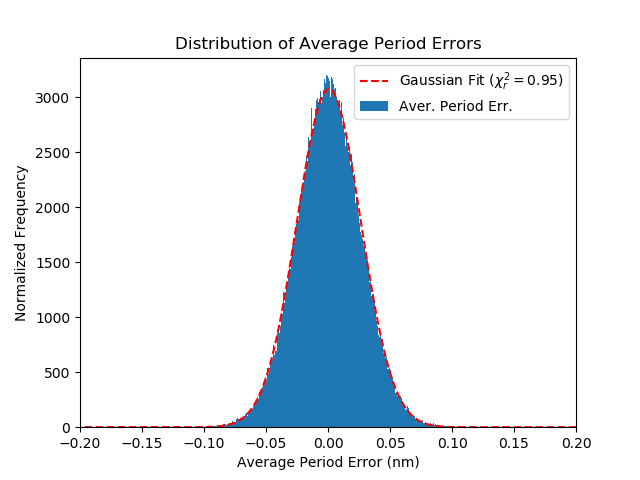}
	\caption{\label{fig:groove_period_distribution} A histogram of the average period errors in Fig. \ref{fig:groove_period_map}A. A Gaussian fit to this distribution is shown as a red dashed line.}
\end{figure}

From the groove displacement map, the average period error as a function of position is derived using Eq. \ref{eq:average_period_error_1d}. Similar to Fig. \ref{fig:groove_displacement_map}, the left side Fig. \ref{fig:groove_period_map} shows the average period error over the grating surface, while, on the right, the average period errors at fixed positions along the groove dimension are shown. To examine the distribution of the average period error over the entire grating, we construct a histogram of the average period error from the 1.2 $\times$ 10$^6$ pixels sampling the grating (Fig. \ref{fig:groove_period_distribution}). We assume Gehrels errors in each bin (\cite{Gehrels_1986}), and fit the resulting distribution with \emph{lmfit}\footnote{\url{https://lmfit.github.io/lmfit-py/index.html}}, a Python-based curve-fitting package \citep{lmfit_2019}. We find that the groove period errors are satisfactorily described ($\chi^2_r$ = 0.95) by a Gaussian distribution. 

We calculate the RMS of the raw average period error map shown in Fig. \ref{fig:groove_period_map}A to be 0.028 nm. This establishes the correct order of a typical period error over the EBL-written grating; however, it should be both corrected for the impact of the ITF and contextualized relative to the interferometer's measurement error in order to provide a faithful assessment of the groove placement accuracy of EBL.

\subsection{Frequency Content of Average Period Error}
\label{subsec:period_frequency}

While the groove period errors are found to be distributed as a uniform Gaussian in magnitude, there is a clear spatial correlation present in Fig. \ref{fig:groove_period_map}A. To elucidate this spatial correlation, we compute the power spectral density (PSD) of the period error map in the dispersion direction by averaging the one-dimensional PSDs of each row i.e., at a fixed groove direction position. As a cross-check on our representative PSD, we calculate the integral of this representative PSD over frequency and find that it agrees with the standard deviation of the groove period error map to within 5\% error, as expected via Parseval's theorem.

\begin{figure}[h]
	\centering\includegraphics[width=\textwidth,trim={0.0in 0.0in 0.0in 0.0in},clip]{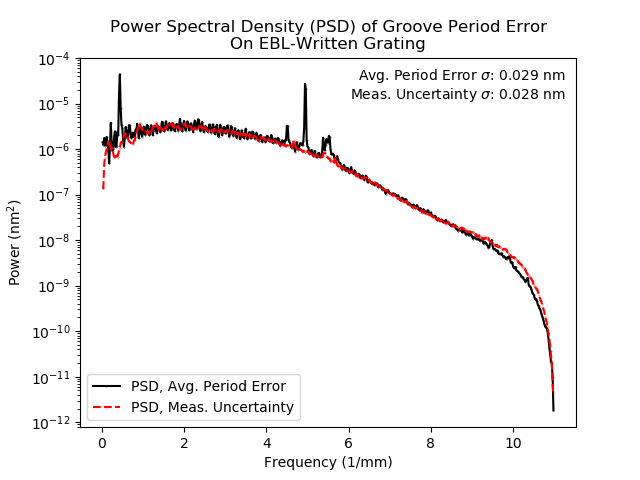}
	\caption{\label{fig:groove_period_psd} The PSD of the average groove period error (Fig. \ref{fig:groove_period_map}A, solid black line) and of the error in the average groove period error (dashed red line), as estimated by propagating the interferometer repeatability via Eqs. \ref{eq:displacement_from_orders} and \ref{eq:average_period_error_1d}. } 
\end{figure}

We next correct this representative PSD with the best-fit ITF from the interferometer calibration (Fig. \ref{fig:itf}, red line); this correction results in a $<$ 5\% change in $\sigma$ (as calculated by integrating the PSD) relative to the uncorrected PSD. The ITF-corrected representative PSD is shown as a black line in Fig. \ref{fig:groove_period_psd}. The average period error $\sigma_d$ as calculated from this PSD is 0.029 nm. Contributing to this average period error is two primary peaks at 0.44 $\pm$ 0.02 cycles/mm and 4.93 $\pm$ 0.02 cycles/mm, which correspond to spatial periods of 2.27 mm and 0.203 mm respectively. 

As a point of comparison, we also compute the PSD of the interferometer error, adopting the measurement repeatability as described in Sec. \ref{subsec:intro_measurements} as the error in $\mathcal{H}_{\pm1}$ and propagating this error through Eqs. \ref{eq:displacement_from_orders} and \ref{eq:average_period_error_1d}. The resulting PSD from the estimated error is displayed as a red dashed line in Fig. \ref{fig:groove_period_psd}. Note that we do not correct this error PSD with the ITF, as this is an estimate of the random error inherent in the measurement as opposed to the systematic underprediction of power from the imperfect ITF. We use the error PSD to estimate the ultimate sensitivity of the interferometric measurements described in Sec. \ref{subsec:interferometer_method} by integrating this PSD to calculate a limiting average period error $\sigma_{d,noise}$. Using the interferometer noise floor in this way results in $\sigma_{d,noise}$ of 0.028 nm.

\section{Discussion}
\label{sec:discussion}

\subsection{Implications for Astronomical Spectroscopy Missions}

The average groove period error calculated in Sec. \ref{sec:results} has implications for the adoption of EBL-written gratings by future astronomical spectroscopy missions. Most science applications for grating spectroscopy, such as line detection or assessing velocity fields, have a figure of merit dependent on the spectral resolution $R$ of an instrument, where $R = \lambda/\Delta\lambda$ and $\Delta\lambda$ is the full width half maximum (FWHM) of the wavelength uncertainty. Following this use of FWHM in the definition of $R$, we adopt $\Delta$$d$ = $2.35\sigma_d = $ 0.068 nm. The factor of 2.35 is derived by relating $\sigma$ and FWHM for a Gaussian; we assert that the assumption of this relationship is well-justified given the Gaussian distribution of period errors shown in Fig. \ref{fig:groove_period_distribution}. 

The resulting estimate of $R$ for the EBL-written grating tested here is $R = d/\Delta$$d =$ 14,600. However, there are two important caveats to this resolution estimate. First, we note that this resolution estimate is dependent on the assumed period $d$. In general, the period of a grating can be assessed in two ways: (1) the average feature size can be measured with nanometer-scale metrology, such as an AFM or SEM, or (2) under illumination by a pencil beam of known wavelength, the relative angle between diffracted and reflected light can be measured precisely and $d$ calculated by use of Eq. \ref{eq:grating_equation}. However, in both instances, the grating period is assessed locally i.e., in a small area as compared to the entirety of the grating. Hence, this period may not be representative of the average period over the entirety of the grating, as assumed for the calculation of the average groove period error in Sec. \ref{subsec:displacement_period_error}. However, recognizing that the overall dependence of $R$ on $d$ is merely $d^{-1}$, we note that a percent uncertainty in $d$ directly maps to a percent uncertainty in $R$. We thus address this quandary of not knowing the true value of $d$ recognizing that, as most often $d$ is known to a few percent, the $R$ reported here should also not be taken to be more accurate than a few percent.  

The second, arguably more important, caveat is the issue of the uncertainty inherent in the interferometric measurement. The reported average period error contains contributions from both the true period error of the grating and the interferometer noise floor. As illustrated in Fig. \ref{fig:groove_period_psd}, the PSD of the interferometric measurement uncertainty is comparable to the PSD of the average period error itself. We therefore argue that the contribution of the interferometer noise floor dominates over the true period error of the grating, and that the estimated limiting spectral resolution $R = $ 14,600 of this grating should be treated as a lower bound for comparable EBL-written gratings. 

In a mission context, period error should be handled as only one component of a comprehensive error budget. Nonetheless, our findings indicate that EBL-written gratings are suitable for missions such as \emph{Arcus} and \emph{Lynx} given the target spectral resolutions of $R > $ 2,500 (\cite{Smith_SPIE_2019}) and $R > $ 5,000 respectively (\cite{Gaskin_2019}).

\subsection{EBL Error Contributions at Specific Frequencies}

Only two frequency components, 0.44 $\pm$ 0.02 cycles/mm and 4.93 $\pm$ 0.02 cycles/mm, have significant power above the measured uncertainty PSD. The error quoted for these frequency components is derived from the width of the component in frequency space. 

These frequency components occur at spatial scales linked to attributes of the EBL-patterning processes. The high frequency component at 4.93 cycles/mm roughly corresponds to the spatial scale of the write field selected for this patterning run, 200 $\upmu$m. While the measured spatial scale does not correspond directly to the scale of this write field to within 3$\sigma$ error (4.93 $\pm$ 0.02 cycles/mm measured vs. 5.00 cycles/mm expected), the measured frequency is dependent on the interferometer pixel scale calibrated at the time of measurement. A systematic error of 1.2\% in this pixel calibration scale is sufficient to explain the difference between the as-measured frequency components and the expected frequencies based on the EBL write, and is a reasonable magnitude given the calibration process.   

Similarly, the low frequency component is in keeping with the frequency of a calibration step performed during the EBL-writing process. For this particular write, the EBL tool alignment is checked every 7 minutes during the course of the write. During this alignment check, the EBL tool translates to fixed alignment markers patterned on the substrate outside of the grating area and recalibrates its position based on these markers.  Examining the tool log, we find that the average separation in the dispersion direction between these calibration points is 2.34 $\pm$ 0.09 mm, in agreement with the measured low frequency component.

In both of these alignment processes during the EBL write, the relationship between the sample position and the (fixed) electron column is adjusted. This adjustment introduces stitching errors, or errors in the position of one write field / calibration field to the next. In the context of a grating, these stitching errors would shift grooves contained in one field relative to another. Thus, EBL stitching errors would be expected to result in a discontinuity in groove placement, and appear as a period error occurring at a fixed spatial scale as observed.

To estimate the contribution of each type of stitching error to the average period error, we integrate the measured PSD around the corresponding frequency peaks. The integral from 0.40 - 0.50 cycles/mm, which encompasses the low frequency peak, contributes $\sigma_{d,f1}$ = 0.009 nm to the overall power. The high frequency peak contributes $\sigma_{d,f2}$ = 0.008 nm to the total measured power, as estimated by integrating the PSD over the range of 4.8 - 5.0 cycles/mm. Adding these in quadrature, a total of $\sigma_{d,EBL} = \sqrt{\sigma_{d,f1}^2 + \sigma_{d,f2}^2} = $ 0.012 nm of the measured power is directly attributable to the EBL-writing process itself. We note that no effort has yet been made to minimize the impact of these calibration steps during the EBL writing process, and hence it may be possible to reduce their impact on $\sigma_d$.

Adopting $\sigma_{d,EBL}$ as a best-estimate of the groove period error inherent in the EBL writing process yields a limiting spectral resolution of $R \sim$ 35,000. In an astronomical context, this spectral resolution is consistent with the medium-resolution echelle spectroscopy capability of Hubble Space Telescope's Space Telescope Imaging Spectrograph (STIS, \cite{Woodgate_1998}), albeit with a finer groove period and without the benefit of the STIS cross-disperser.

\subsection{Future Measurements of Curved Substrates}

An important open question is whether the EBL groove placement accuracy as quantified here will translate faithfully to the patterning of more complex grating patterns, e.g., gratings with arbitrary groove orientations or freeform surfaces. Such gratings break the simplifying symmetry of the interferometric measurement technique for constant period, flat gratings as presented in Sec. \ref{subsec:interferometer_method}. In principle, however, the interferometric technique can be adapted to assess the expected groove placement accuracy on a customized EBL grating by writing a grating pattern yielding a back-diffracted wavefront when illuminated by a plane wave.

To perform such a measurement, a test substrate with a figure representative of the desired customized grating is needed. Assuming a fixed incidence geometry on this test substrate, the local groove density and orientation yielding a back-diffracting wavefront can be calculated numerically via a raytracing program. This back-diffracting grating pattern would then be written on a test substrate, and the back-diffracting wavefront measured interferometrically. Additionally, detailed knowledge of the figure of the test substrate is required, either from manufacturing metrology or via measurement with a computer-generated hologram. This independent measurement of the test substrate can be used to account for the phase errors introduced in a back-diffracted wavefront due to figure. Subtracting the impact of the test substrate figure from the back-diffraction interferogram yields a residual phase error due to imperfect groove placement on this curved substrate. 

\begin{figure}[h]
	\centering\includegraphics[width=3.5in,trim={0.0in 0.0in 0.0in 0.0in},clip]{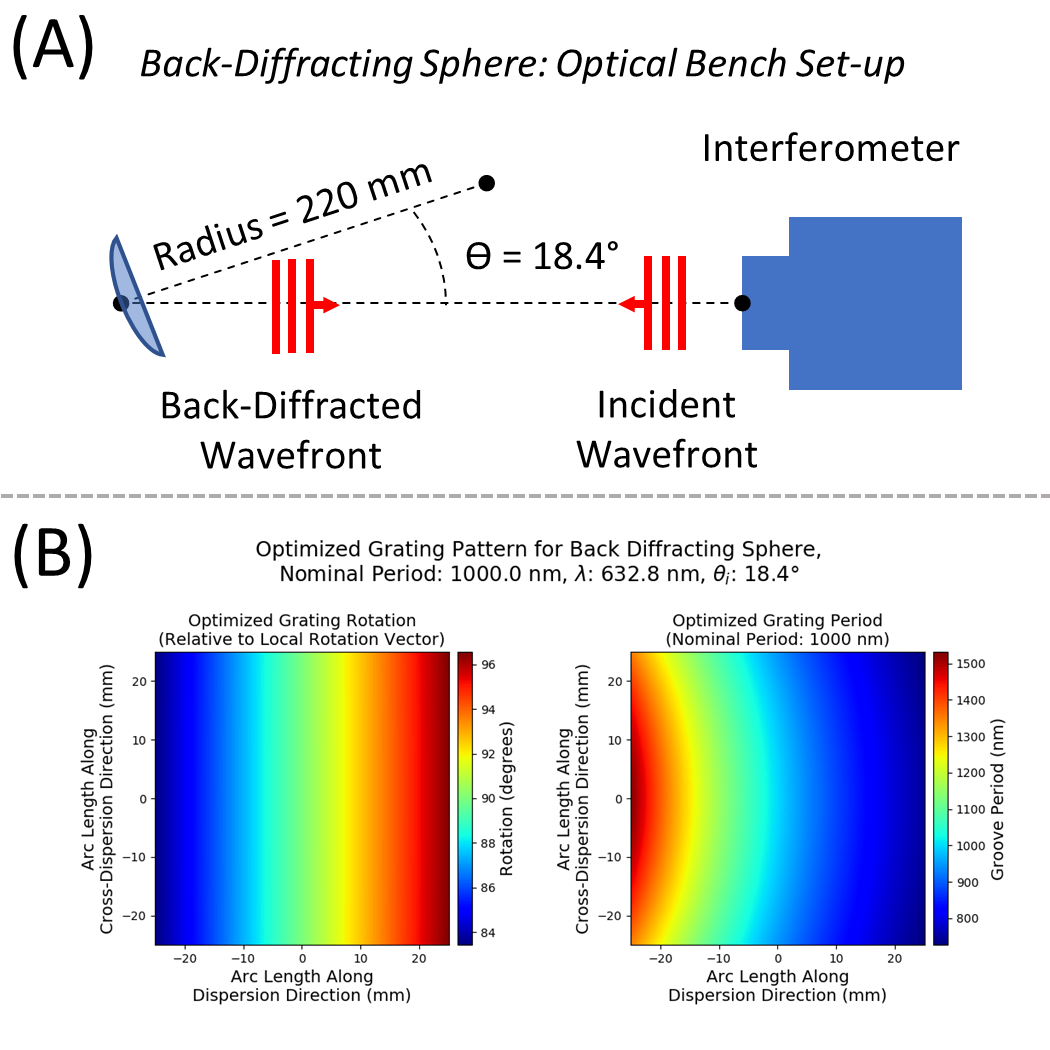}
	\caption{\label{fig:backdiffracting_sphere} (A) A diagram showing the optical bench configuration for assessing the groove displacement of a spherical EBL grating. (B) The grating groove orientation (left) and period map (right) required to yield a back-diffracting sphere. Errors in realizing this pattern with EBL lithography will result in phase errors in the interferogram of the back-diffracted order.}
\end{figure}

As a concrete example, we have calculated the customized grating pattern required to assess the groove placement accuracy on a spherical grating. The assumed substrate and incidence geometry for this test grating is shown in Fig. \ref{fig:backdiffracting_sphere}A. The sphere is assumed to have a diameter of 76.2 mm and a radius of curvature of 220 mm. The sphere would be patterned with a grating over at least the central 50 mm square and illuminated by a plane wave of the same size. The groove direction vector and grating period required to yield a back-diffracted wavefront assuming this substrate and incidence geometry are shown as the lefthand and righthand plots in Fig. \ref{fig:backdiffracting_sphere}B. Assessing the groove placement accuracy of EBL on a curved substrate such as this simple sphere would serve as a proof-of-concept for the customized gratings discussed in Sec. \ref{sec:intro}, and reduce the risk of adoption for instrument concepts featuring these gratings such as the two-element spectrometer (\cite{DeRoo_SPIE_CDXO_2019}). 

%
\section{Conclusions}
We have fabricated a large-format, 1000 nm period grating on a relatively thin (1.5mm) Si substrate, and assessed its limiting spectral resolution in the context of astronomical spectroscopy missions. Our core findings are as follows:

\begin{itemize}
\item{The groove placement accuracy and average period error $\sigma_d$ over the grating surface have been calculated from interferometric measurements of the $\pm$1st orders. Correcting for the interferometer ITF, we find $\sigma_d$ = 0.029 nm.}
\item{Using only the $\pm$1st diffracted orders, the figure of the grating as measured in 0th order was reproduced to within the repeatability error of the interferometer. This method serves as a cross-check on the interferometric approach to assessing grating performance and the self-consistency of the measurements employed in this work.}
\item{We find that the measured distribution of period errors on this EBL-written grating to be well-described by a Gaussian distribution.}
\item{Based on a representative PSD, we identify two frequency components associated with the stitching error of the EBL fabrication process. These frequency components contribute $\sigma_{d,f1}$ = 0.009 nm and  $\sigma_{d,f2}$ = 0.008 nm to the overall power of the average period error.}
\item{Outside of these two EBL-related frequency components, the PSD is similar in shape and magnitude to the PSD expected from propagating the interferometer measurement error. Moreover, the overall power attributable to the error is 0.028 nm.}
\item{We estimate the limiting spectral resolution of this EBL-written grating as $R \sim$ $d/\Delta$$d$, where $\Delta$$d$ is the FWHM of the average period error. This yields $R = $ 14,600 for EBL-written gratings. We argue that, given the contributed power of the interferometric measurement error, this should be interpreted as a lower bound. Performing the same assessment for the stitching error features with power above the interferometer noise floor yields $R = $ 35,000.}
\end{itemize} 

Based on these measurements, we conclude that, in principle, EBL-written gratings are a suitable grating technology for spectroscopy missions requiring periods $\sim$1000 nm and spectral resolutions of $R > $ 10,000, and may be capable of supporting missions of $R > $ 30,000. Additional improvements to the fabrication process are likely needed to support instruments with resolutions of $R > $ 100,000, such as the POLLUX instrument for the LUVOIR concept. As an aside, we also note that, since limiting average period error $\sigma_{d,noise}$ scales with $d$ and is of order a few tenths of an Angstrom for a grating with period $d$ = 1000 nm, improvements to interferometric measurement techniques may also be needed to support the testing and calibration of gratings for systems $R > $ 100,000. 

Finally, the fidelity of EBL patterning can be assessed for gratings on curved substrates by adapting the interferometric technique employed here. Such a study would be an important technical milestone towards realizing custom EBL-written gratings and enabling future innovative spectroscopy missions.   

\section*{Acknowledgments}
This work was supported by NASA grants 80NSSC19K0661. C. DeRoo acknowledges support from the Early Career Investigator Research Program of the Iowa Space Grant Consortium (ISGC). J. Termini's efforts were partially supported by the ISGC's Undergraduate Research Award. B. Donovan is supported by a NASA Space Technology Research Fellowship (80NSSC18K1178). Part of this work was carried out in the Nanofabrication Laboratory at Penn State's Materials Research Institute. This work makes use of \emph{PyXFocus}, an open source Python-based raytracing package.

%

\bibliography{ebl_backdiffraction_references}
\bibliographystyle{aasjournal}


\listofchanges

\end{document}